\documentclass{aa}
\usepackage{graphicx}
\usepackage{multirow}
\usepackage{txfonts}
\usepackage{floatflt}
\usepackage[authoryear]{natbib}
\bibpunct{(}{)}{;}{a}{}{,}

\begin{document}
\title{Evolution of the progenitor binary of V1309 Scorpii before merger}
\author{K. St\c epie\'n}

\institute{Warsaw University Observatory, Al.~Ujazdowskie~4, 
          00-478~Warsaw, Poland, email: kst@astrouw.edu.pl}
\date{Received; accepted}
\abstract
{It was recently demonstrated that the eruption of V1309 Sco was a result
of a merger of the components of a cool contact binary.} 
{We computed a set of evolutionary models of the detached binaries 
with different initial parameters to compare it with
pre-burst observations of V1309 Sco.}
 {The models are based on our recently developed
evolutionary model of the formation of cool contact binaries.}
 {The best agreement with
observations was obtained for binaries with initial masses of 
1.8-2.0 $M_{\sun}$ and initial periods of 2.5-3.1 d. The evolution of
these binaries consists of three phases: at first the
binary is detached and both components lose mass and angular
momentum through a magnetized wind. This takes almost two thirds of the total
evolutionary lifetime. The remaining third is spent 
in a semi-detached
configuration of the Algol-type, following the Roche-lobe overflow by the
initially more massive component. When the other component leaves the main
sequence and moves toward the giant branch, a contact configuration is
formed for a short time, followed by the coalescence of both components.} 
{}
\keywords{stars: individual: V1309 Sco - stars: binaries: close - stars:
  late type - stars: evolution
} 

\titlerunning{Progenitor of V1309 Sco}
\authorrunning{K. St\c epie\'n}
\maketitle
\section{Introduction}

V1309~Sco erupted in 2008, which increased its brightness by about 10
magnitudes. The discussion of spectral and photometric observations
obtained after the eruption showed a close resemblance of the outburst of
V1309~Sco to that of V838 Mon and outbursts of a few other variables
classified as ``red novae'' \citep{mas10}. Several explanations of these
unusual eruptions have been offered \citep[][and references
therein]{mas10,tyl11}.  In particular, \citet{st03} and \citet{ts06}
suggested that a collision of two stars (or a star and a substellar object)
best reproduces the observed characteristics of the red novae outbursts,
however, till recently, other mechanisms were considered as well.  Fortuitously,
V1309~Sco lies in one of the stellar fields observed by the
OGLE team, so several years of precise photometric monitoring preceding the
eruption of the variable exist. These data are analyzed by
\citet{tyl11}. The observations clearly show that the progenitor of
V1309~Sco was an eclipsing contact binary with a period of about 1.4 d,
which rapidly shortened its period and substantially changed the light
curve just before the eruption. The observations obtained after the
outburst show no trace of periodicity. That leaves no doubt that the
outburst of V1309~Sco was the result of a violent collision of the binary
components.

The purpose of the present paper is to model the evolution of the
progenitor of V1309~Sco from zero age main sequence (ZAMS) to the
eruption.     

\section{Model}
\subsection{Description of the model}

Recently we developed new model of the formation of cool contact binaries 
\citep{ste04,ste06a,ste06b,ste09}. The
model assumes that contact binaries emerge from cool detached
binaries with initial orbital periods of a few days. Both components
possess subphotospheric convective zones that are necessary for generating
surface magnetic fields. This limits the mass of each component to
values less than or equal to 1.3 $M_{\sun}$. For more massive stars the coronal
X-ray surface flux decreases so rapidly with increasing mass
\citep{schm97} that they are not expected to possess significant
magnetized winds, which originate in hot coronae. The magnetic activity level
of a cool star increases with the increase of its rotational angular
velocity \citep[e.\thinspace g.][]{piz03}. Assuming the spin-orbit 
synchronization of rotation, the activity of both components of a
short-period binary is very high, at the so-called ``saturation'' level.
The magnetic activity drives stellar winds, carrying away mass and spin
angular momentum (AM). Because of synchronization, the AM loss (AML)
results in a decrease of the orbital AM, a shortening of the orbital
period, and an actual spin-up of both components.

Neglecting the spin AM of the components we have

\begin{equation}
H_{\mathrm{orb}} = 1.24\times 10^{52}M^{5/3}P^{1/3}q(1+q)^{-2}\,.
\label{horb}
\end{equation}

The AML rate of a close binary is given by \citep{gs08}

\begin{equation}
\frac{{\mathrm d}H_{\mathrm{orb}}}{{\mathrm d}t} = -4.9\times 10^{41}
(R_1^2M_1 +R_2^2M_2)/P\,.
\label{dhdt}
\end{equation}

Here $H_{\mathrm{orb}}$ is orbital AM in cgs units, $P$ - period in days,
$M_{1,2}$ and $R_{1,2}$ are masses and radii of the components in solar
units and $t$ is time in years.  The formula is based on the semi-empirically
determined AML rate of single, cool stars \citep{ste06b}. The uncertainty
of the numerical coefficient is about 30\%.

The mass loss rate due to the wind at its highest, saturated strength,
adopted here for each
component, is based on the empirical determination by \citet{wood2002}

\begin{equation}
\dot M_{1,2} = -10^{-11}R_{1,2}^2\,,
\label{dmdt}
\end{equation}

where mass loss rates are in $M_{\sun}$/year and radii in solar units.
This relation applies to stars with $M_{1,2} \le 1 M_{\sun}$. For
  stars with higher masses $R_{1,2} \equiv 1$ is adopted.

Mass loss rates of cool MS stars belong to the least known parameters. The
values from $1.3\times 10^{-10} M_{\sun}$/year \citep{dem06} down to
$2\times 10^{-13} M_{\sun}$/year \citep{holz07} can be found in the literature.
Eq.~(3) results from a relation between the directly observed mass loss
rate {\it per unit surface area} and X-ray surface flux, see Fig.~7 in
\citet{wood2002} and Fig.~3 in \citet{wood2005}. The latter authors
published measurements of three additional solar type stars. They included
70 Oph with a very high value of the mass loss rate into a final fit, but
rejected $\xi$ Boo, considering it as an outlier whose value
of this rate is too low. If, however, both stars are included, one obtains the
relation $\log\dot m \propto (1.01\pm 0.24)\log F_x$, where $\dot m$ is the
stellar mass loss per unit surface area, expressed in the units of the
solar mass loss rate. Deviations of 70 Oph and $\xi$ Boo from the line
described by this relation are roughly the same. The relation suggests a
strict proportionality of mass loss rate and X-ray flux, both per unit
surface area. For the saturated state $F_x = 1-2\times 10^7$ (in cgs
units), hence $\dot m \approx 300-700\dot m({\rm Sun})$. We adopt 500 for
this coefficient and with $\dot M({\rm Sun}) = 2\times 10^{-14}
M_{\sun}$/year we obtain Eq.~(3). The estimated uncertainty of the 
numerical factor in it is of order a factor of two.

Two more, albeit indirect, estimates give similar values for the mass loss
  rate. From a linear relation between mass loss rate and X-ray
  flux an upper limit for this rate can be found, applying the so-called,
  Reimers formula to rapidly rotating, solar type stars. The value 
$1.5\times 10^{-11} M_{\sun}$/year is obtained
  \citep{led08}. Another rough estimate of the mass loss rate of these
  stars results from an observation that for magnetized winds the
  Alfv\'en radius is close to 10 stellar radii and varies little from star
  to star \citep{shore92}. Taking Eq.~(2) for one component with solar
  values of parameters and remembering that AML rate equals to
  $(R_A)^2\omega\dot M$, where $R_A$ is the Alfv\'en radius and $\omega$ -
  stellar angular velocity, we obtain after simple calculations, $\dot M =
  0.8\times 10^{-11} M_{\sun}$/year. Both results are close to the
  value appearing in Eq.~(3).

We assign the subscript ``1'' to the initially {\em more} massive stars and
keep it this way also after mass ratio reversal, when this subscript
denotes the {\em less} massive star. The mass ratio $q \equiv M_1/M_2$ so
initially $q > 1$ and after the mass reversal $q < 1$.

The equations are combined with the third Kepler law and the approximations
for Roche lobe sizes given by \citet{eggl83}. The set of equations is
integrated in time to follow the evolution of binaries with various initial
masses and periods. We assume that the evolution of internal structure
of each component can be approximated by the single star evolution. Models
obtained by \citet{gira00} and a set of low-mass models calculated for us by
Dr. Sienkiewicz were used.\footnote{A description of
the models is given in \citet{ste06a} and the details of the modeling
program are given in \citet{pacz07}.} We also assume that both
components independently lose AM through the magnetized wind, i.\thinspace e. any
interaction between them is neglected. The calculations start when the
binary is on the ZAMS. It is detached at that time and AML results in shrinking
of the orbit accompanied by the slow mass loss of the components. At the
same time the components expand because of their internal evolution. At one
point the shrinking Roche lobe of star ``1'' descends onto its surface and
the (first) Roche lobe overflow (RLOF) occurs. This is followed by fast
mass transfer from star ``1'' to star ``2''. This process has not yet been
satisfactorily modeled. The existing models suggest that the rapid
transfer of a small amount of mass results in an expansion of star ``2''
followed by filling of its Roche lobe \citep{webb76}. A contact binary is
formed. Following the idea put forward by \citet{lucy68,lucy76} and
\citet{fla76}, many authors assume that the expanded component does not
relax thermally and stays oversized, transferring mass back to its more
massive companion on the evolutionary time scale. It was, however,
demonstrated by \citet{ste09} that the energy transfer between components
of a contact binary takes place in the form of a large-scale circulation
bound to the equatorial region of the less massive component. The core
energy of that component can be freely radiated by the polar regions and
the star remains in thermal equilibrium. This result indicates that star
``2'' will relax thermally when the fast mass transfer is stopped. In other
words, it is assumed that mass transfer in cool close binaries is
conceptually similar to the process taking place during the formation of
semi-detached Algols. This process results in mass ratio reversal. Some
amount of mass and AM may likely be lost during the common envelope
phase but without reliable models of this phase, two additional free
parameters would have to be introduced to allow for the losses. To avoid
that, we assume conservative mass transfer during the RLOF -- an assumption
commonly accepted when modeling the formation of Algols.
 
For $P_{\mathrm{init}} \approx 2$ d the time scale for AML is roughly equal
to the evolutionary time scale of star ``1'' \citep{ste06a,gs08} and varies
with period as $P^{4/3}$. Shorter initial periods result in early mass
transfer when both components are still on the MS, whereas in binaries with
initial periods exceeding 2 d mass transfer takes place when star ``1'' is
depleted of hydrogen in the center or even possesses a small helium
core. After the fast mass exchange the mass continues to
flow from star ``1'' to star ``2'' at the much lower rate owing to the slow
evolutionary expansion of star ``1'' accompanied by AML through the magnetized
wind. Mass transfer lengthens the orbital period, whereas AML
acts in the opposite direction. Time evolution of the orbital period and mass
ratio is governed by interplay between these two processes. The
instantaneous mass transfer is usually computed from a parametric formula
\citep[e.~\thinspace g.][]{eggl02}, relating it to the amount by which the star
overflows its Roche lobe. Instead of calculating that rate at each time
step, an average (constant) mass transfer rate can be used. Its value
results from the condition that the star fits into the Roche lobe at the
beginning and at the end of the considered time interval \citep{gs08}. This
is the approach we applied here.

Note that the above model does not contain any free parameters. The only
adjustable parameter is the average mass transfer rate in the phase of slow
mass transfer past fast mass exchange. Once the value of this parameter is
found, the model is fully determined, which means that a binary with
specified values of the initial orbital period and component masses evolves
up to the final stage along a unique track.

\subsection{Observational constraints and evolutionary scenario}

The main constraints on the parameters of the progenitor of V1309
result from its photometric observations prior to the outburst
\citep{tyl11}. The first useful observations from 2002 show a regular light
curve corresponding to a cool contact binary with an orbital period of
1.438 d. From two-color observations
\citet{tyl11} estimate the effective temperature of the binary at 4500
K. The uncertainty of this value can be estimated at 500 K. 

Based on the red color of the binary, massive hot components are
excluded. Because the components fill their Roche lobes,
they must both be substantially oversized with respect to their expected MS
sizes. This is true for any mass ratio and consequently both stars must be
evolutionary advanced. 

The observational constraints require that the initial orbital period
should be
longer than 2 days and the initial mass of star ``1'' should be higher than 1
$M_{\sun}$, because this component must complete its MS evolution within
the age of the binary. Indeed, its MS life time should be appropriately
shorter than that because star ``2'' also needs time to complete its MS
evolution prior to the eruption. The latter condition means that the initial
mass of star ``2'' cannot be too low. 

We assume that the outburst was
triggered by an instability resulting from a rapid increase of the
spin AM of the more massive component above the limiting value of about a third
of the orbital spin AM \citep{rs95,eggl01}, following its evolutionary
expansion and filling the Roche lobe. The transfer of orbital AM into spin AM
brings both components closer to one another so that they overflow
the outer critical Roche surface and mass and AM are lost very efficiently
through the L$_{\mathrm 2}$ point \citep{webb76}. The process results in a rapid merging
of the components. In contact binaries the
instability sets in when the mass ratio reaches a critical value $q \approx
0.07-0.10$ (the exact value depends on the detailed structure of the
binary).

\begin{table}
\caption{Main properties of the evolutionary models of the possible
  progenitors of V1309 Sco}
\label{tab}
\centering 
\begin{tabular}{lccrc}
\hline
\hline
 ev. stage & age & stell. masses & Orb. AM & period \\
& Gyr & $M_{\sun}$ &  $\times 10^{51}$ & days \\
\hline
 initial (ZAMS) & 0 & 1.10+0.90 & 13.22  &  2.50 \\  
 start RLOF & 8.08 & 1.01+0.85 &  8.65 &  1.00 \\
 end mass ex. & 8.57 & 0.50+1.33 & 7.70 &  1.46 \\
 sec. RLOF & 11.4 & 0.17+1.54 &  3.12 & 1.42 \\
\hline
 initial (ZAMS) & 0 & 1.20+0.60 &  10.34 &  2.80 \\
 start RLOF & 5.32 & 1.15+0.58 &  6.89 &  1.00 \\
 end mass ex. & 5.45 & 0.51+1.22 & 6.77 & 1.20 \\
 sec. RLOF & 7.85  & 0.16+1.52 &  2.81 &  1.43 \\
\hline
 initial (ZAMS) & 0 & 1.30+0.50 &  9.66 &  3.10 \\
 start RLOF & 4.16 & 1.26+0.49 &  6.59 &  1.12 \\
 end mass ex. & 4.33 & 0.40+1.35 & 6.40 & 1.57 \\
 sec. RLOF & 7.01  & 0.16+1.54 &  2.81 &  1.44 \\
\hline
\multicolumn{4}{c} {Mass loss rate reduced by a factor of 5} \\
\hline
 initial (ZAMS) & 0 & 1.10+0.90 & 13.89  &  2.90 \\  
 start RLOF & 7.68 & 1.08+0.89 &  9.24 &  0.93 \\
 end mass ex. & 8.23 & 0.53+1.43 & 8.64 &  1.49 \\
 sec. RLOF & 10.2 & 0.26+1.69 &  4.81 & 1.42 \\
\end{tabular} 
\end{table}

\begin{figure}
\includegraphics[height=\hsize]{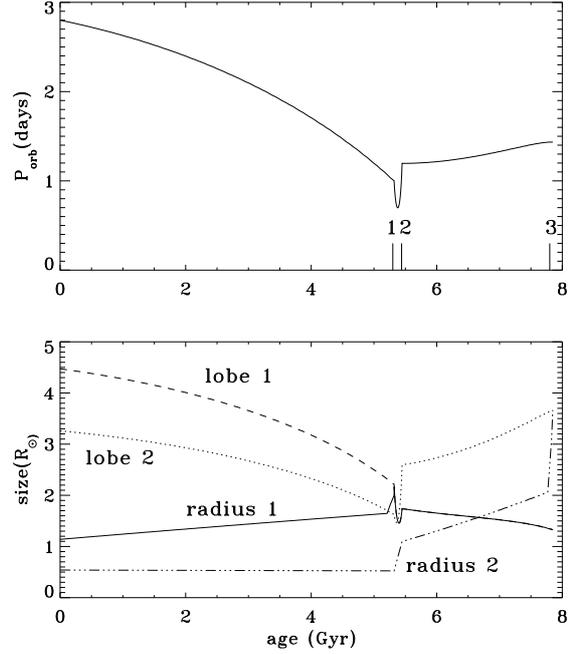}
\caption{Variations in time of the orbital period (top) and geometrical
  parameters of the binary with initial masses 1.2+0.6 $M_{\sun}$ and
  initial period of 2.8 d. Short vertical bars denote: 1 - beginning of the
  fast mass transfer phase from star ``1'' to star ``2'', 
  2 - end of the fast mass transfer phase, 3 - rapid increase of radius and 
  moment of inertia of star ``2'' resulting in instability.}
\label{bin12}
\end{figure}

We calculated the evolution of binaries with initial masses between 1.1 and 1.3 $M_{\sun}$
for star ``1'' and the initial masses between 0.5 and 1.1 for star ``2''
from the ZAMS to the pre-burst stage. We did not consider
binaries with two identical components ($q_{\mathrm{init}} = 1$).

\section{Results and conclusions}

As the detailed calculations show, very few models from the considered set
lead to the binary with the observed pre-burst properties. The majority of the
considered binaries end their evolution either as W UMa-type stars with
periods of a fraction of a day if they lost too much AM, or as
Algols with periods of several days if they lost too little AM.  This
divergence of the period is mostly due to the period dependence of the AML
rate (Eq.~\ref{dhdt}): the shorter the period, the higher the AML rate, and
{\it vice versa}. The initial parameters must have just the right values 
so that with a specified AML rate, star ``1'' has enough
time to complete its MS evolution and the mass exchange takes place in the AB
case. The following evolutionary phase of the binary must take just so much
time that star ``2'' completes its MS evolution and migrates to the base of
the red giant branch.

Three models from the considered set fulfill all the requirements and
reproduce the observations of V1309 Sco correctly. Their evolutionary
parameters are given in Table 1 (the first three models). 
Fig.~\ref{bin12} shows the time evolution
of the orbital period (top) and the geometrical parameters (bottom) of one
of the models. The time behavior of all three models is similar, although
the total time scales are different. In particular, the total age of
  the first models exceeds 11 Gyr, which is rather long for a star from the
  galactic disk. Note a rapid increase of the radius of
star ``1'' shortly before phase 1 and of star ``2'' at phase 3. The stellar
radii are plotted very approximately in Fig.~\ref{bin12} between phases 1
and 2 because both stars are out of thermal equilibrium.  In the final stage
both components possess small helium cores with a mass of 0.10-0.15
$M_{\sun}$.

We provide here a more detailed description of the evolutionary models.
The initial period of each binary is long enough that star ``1'' reaches
TAMS still in the detached phase but its Roche lobe has a size just
slightly larger than the TAMS radius. The star expands while it builds a small
helium core, and fills its lobe (phase 1 in Fig.~\ref{bin12}). Fast mass
transfer follows. When both components reach thermal equilibrium, an Algol
is formed with a period between 1.2 and 1.6 d (phase 2 in
Fig.~\ref{bin12}). Between phase 2 and 3 mass transfer continues at a
very low rate while the helium core of star ``1'' slowly grows. At the
same time star ``2'' evolves across MS. After a few Gyr it reaches the TAMS and
moves toward the red giant region. It fills its Roche lobe when it is at
the bottom of the red giant branch (phase 3). It possesses a helium core 
at that time
with a mass of about 0.15-0.16 $M_{\sun}$. All three binaries look very
similar just before the eruption: star ``2'' has a mass of about 1.5
$M_{\sun}$ the mass ratio $q \approx 0.1$ and the orbital period is about
1.43 d. These values of the binary parameters, in particular the extreme
mass ratio, were not assumed in the modeling procedure. On the contrary,
they result from the calculations. The mass transfer rate in the
semi-detached phase is uniquely determined by the
rate of evolution of star ``1'' together with the influence of the wind
removing mass and AM from the system. The resulting values of mass transfer
rate were between $0.8-1.3\times 10^{-10}  M_{\sun}$/year, depending on the
binary. Altogether, about 0.3 $M_{\sun}$ were transferred in 2.5-3 Gyr
- the time needed to finish the MS evolution by star ``2'' (between phase 2
and 3 in Fig.~\ref{bin12}).

Table~\ref{tab} shows that V1309 Sco might have been produced
from a binary with the following initial parameters: more and less massive
components from the range of 1.1-1.3 $M_{\sun}$ and 0.9-0.5 $M_{\sun}$,
respectively and the orbital period from the range 2.5-3.1 d with the
additional condition that the total mass should not exceed 2
$M_{\sun}$. The pre-burst contact binary V1309 is different from W UMa-type
stars. These stars have their primaries on the MS and stay in contact
for a long time, on the order of Gyrs \citep{ste06a}. The primary of V1309
Sco is a giant that recently filled its Roche lobe and the contact phase
is very short, ending with coalescence.

The luminosity and the effective temperature of star ``2'' just before the
outburst is equal to 7.5 $L_{\sun}$ and 5000 K, respectively. With this
luminosity the distance to the star is equal to 3700 pc. Assuming that star
``2'' transfers a fraction of its luminosity to star ``1'', which is
observed in contact binaries \citep{ste09}, the effective temperature of the
whole binary is equal to about 4840 K, which is close to the temperature
estimated by \citet{tyl11}. If the eruption resulted in merging of both
components, the post-eruption star has a mass of about 1.7 $M_{\sun}$
with a helium core of about 0.2-0.3 $M_{\sun}$. The star will relax to the red
giant stage.

The instability resulting in the merging of both components was triggered by a
dramatic increase of the moment of inertia of star ``2'' when it approached
the base of the red giant branch. It increased by a factor of several, up
to the value of $1.8\times 10^{55}$ in cgs units \citep{rut88}. The
migration time to the red giant region is on the order of $10^8$ years,
which is slow enough to keep the synchronization of its rotation with
orbital period. As a result the spin AM of star ``2'' reaches $9\times
10^{50}$ in cgs units compared to about $3\times 10^{51}$ for orbital AM. 
The ratio of the spin AM to the orbital
AM approaches 1/3, which is a sufficient condition for the occurrence of
instability.

The uncertainties of the computed models come mostly from the uncertainty
of the coefficients in Eqs.~(\ref{dhdt}-\ref{dmdt}) and the accuracy of the
evolutionary models of both components. An increase/decrease of the
efficiency of AML via the wind by 30\% requires an increase/decrease of the
initial period by about 0.2 d to preserve the required amount of AM in the
final, pre-burst model. A change of the mass loss rate by a factor of two
requires a corresponding change of the initial masses by about 0.1
$M_{\sun}$. Of particular importance is the case of a lower mass loss rate
than resulting from Eq.~(3). Let us take an example of the first model in
Table~1. For the mass loss rate decreased by a factor of two, the initial
mass of each component would have to be decreased by 0.1 $M_{\sun}$ to
obtain nearly identical pre-burst model as listed at the top of
Table~1. However, the total age of this model would be higher by about 10
\%. To see the effect of a {\em substantial} decrease of the mass loss rate
(beyond our estimated uncertainty), a series of models with the same
initial masses as the first one in Table~1 were computed but with different
initial orbital periods. The employed mass loss rate was five times lower
than resulting from Eq.~(3) i.\,e. at the level of the highest directly
measured rate \citep{wood2005}. The model that best reproduced the
pre-burst binary was selected and is shown at the bottom of Table~1. The
reduced mass loss rate influences the final model in several ways. The
masses of both components are higher, which shortens the evolutionary time
scales. A low mass loss rate accelerates the orbit tightening compared to
the high rate. This is why the initial period had to be lengthened to a
value 2.9 d, compared to 2.5 d in the first model in Table~1. Moreover, the
final mass ratio has not reached a critical value for the Darwin
instability to occur. Nevertheless, after star ``2'' left MS and approached
the giant branch, a common envelope should develop, which may also result
in a merging of both components.

An estimate of the uncertainties
connected with the evolutionary models of the components is more
difficult. Even the models of single stars differ significantly among
different authors. In particular, time scales of the consecutive
evolutionary stages and radii of the models may differ by several percent -
compare, e.\,g., \citet{scha92} and \citet{gira00} \citep[see
also][]{torr10}. The differences result from different input physics,
methods of computation, additional effects such as overshooting, the presence
of magnetic fields, etc. The uncertainties resulting from interpolation in
the existing sets of models or from differences between single star models
and interacting binary component models are also difficult to assess. We
assume that all those uncertainties amount to $\pm 10\%$ on the input
parameters and the final parameters of the pre-burst binary. The generous
estimate of all uncertainties indicates that the ZAMS progenitor of V1309
Sco was a binary with the period between 2.3-3.3 d, the total mass between
1.7 and 2.2 $M_{\sun}$ and the mass of the more massive component between
1.1 and 1.3 $M_{\sun}$.

\bibliographystyle{aa}

\end{document}